\documentclass[prl,apsi,twocolumn]{revtex4}

\usepackage{dcolumn}
\usepackage{graphicx}
\usepackage{latexsym}
\usepackage{amsmath}
\usepackage{amssymb}

\newcommand{\be}{\begin{equation}}
\newcommand{\ee}{\end{equation}}
\newcommand{\bea}{\begin{eqnarray}}
\newcommand{\eea}{\end{eqnarray}}
\newcommand{\bfig}{\begin{figure}}
\newcommand{\efig}{\end{figure}}

\newcommand{\sech}{\hbox{sech}}

\begin{document}

\title{\Large \bf Localization of fields on brane}

\author{Ratna Koley${}^{}$ \footnote{E-mail: tprk@iacs.res.in}}
\affiliation{ Department of Theoretical Physics and Centre for
Theoretical Sciences,\\
Indian Association for the Cultivation of Science,\\
Kolkata - 700 032, India}

\begin{abstract}
Warped braneworld scenario provides an alternative to the usual Kaluza-Klein compactification of extra dimensions through the so-called localization of fields. Randall-Sundrum type models do not directly prove, rather, assume that all standard model fields are localized to
the TeV brane. I will provide the mechanism for addressing the problem and discuss the issue of localization of fields on the brane for various types of braneworld models available in the literature.
\end{abstract}

\maketitle

\section{Introduction}

The warped branewolrd model introduced by Randall-Sundrum (RS) \cite{RS} provides a novel solution of the gauge hierarchy problem in particle physics. Also it was suggested by Randall and Sundrum that gravitational interactions between particles on a brane, which separates two patches of $AdS_5$ spacetime, could have the correct four-dimensional Newtonian behaviour, provided that the bulk cosmological constant and the brane tensions are fine tuned \cite{RS}. This idea gave rise to mammoth activities in many interesting fields of physics like particle phenomenology \cite{phenom1,phenom2}, cosmology \cite{cosmo} etc. The generalisation of RS framework to higher dimensions has been also studied in the literature \cite{higher}. The possible experimental probes also has been studied extensively \cite{expprobe}. The consequences of the presence of bulk fields have also been explored rigorously \cite{bulkfield}. The crucial ingredient of this scenario is a hypersurface (brane) on which the Standard Model (SM) fields are assumed to be localised. This is inspired by string theory. The SM fields are represented by open strings and one of the ends of a string  is attached to the brane. So the SM fields are naturally localized on the D-brane according to string theory. But there is no such strong footing behind the assumption of having SM fields on the barne in RS scenario. Therefore an important issue in the study of braneworld models is the question of localization of the SM fields on the brane. 

An explicit study of the bulk fields in the RS background reveals that the scalar 
(spin 0) and gravity (spin 2) can be localized on the positive tension 
brane \cite{bajc}. It has been shown \cite{fermiloc} that massless fermions in a warped geometry with a RS warp factor cannot be localized on the positive tension visible brane.
On the other hand spin $\frac{1}{2}$ 
and $\frac{3}{2}$ fermions are confined to a brane with negative tension.
 However with additional couplings, say, Yukawa coupling with bulk scalar it is possible to have localized massless chiral fermions on the positive tension visible brane. First such mechanism has been developed by Randjbar-Daemi and Shaposhnikov \cite{fermiloc} in 4D flat spacetimes. But neither the positive tension nor the negative tension branes are capable of localizing gauge fields in the minimal setup \cite{5dloc}.

 In this paper we will present a brief overview of the localisation scenario on the  brane. We will focus mainly on the single brane {\sc{rs-ii}} model where Newtonian gravity is recovered on the visible positive tension brane with a high energy correction due to an infinitely extended extra dimension.
The role of bulk scalar fields on the warped geometry as well as on the localization of fields will also be discussed in greater detail. The organization of the paper is as follows. In section two we discuss the general localization mechanism on the brane. The issue - how the localization scenario differs in different braneworld models - is addressed in the next section. We summarise the results in the last section. 

\section{ Localization Mechanism : in general}

In order to explicitly obtain localization, ordinary matter fields are considered 
to be extended over the full five dimensional bulk. From the behavior of {\sc kk}
modes along the extra dimension we determine whether the fields are
localized on the brane. 

The general metric representing the warped geometry with an extra dimension is given by
\be
\label{metric}
ds^2 = d\sigma^2 +e^{-2f(\sigma)} \eta_{\mu\nu}  dx^{\mu} dx^{\nu}
\ee

where $\sigma$ is the extra spacial dimension, ${\mu, \nu}$ indices stand for brane coordinates. The warping of the geometry is characterised by the factor $f(\sigma)$.
To address the localization issue, one needs to assume SM fields as propagating in the bulk (i.e. SM fields functionally dependent on higher dimensional coordinates). Let us first consider the case of fermions as an example. The five dimensional spinor can be decomposed into four dimensional and fifth dimensional parts: 

\be
\Psi(x^{\mu}, \sigma) = \psi(x^{\mu}) \xi(\sigma)
\ee

The action for  a Dirac fermion propagating in the five-dimensional warped space (\ref{metric}) is :

\be
\int \sqrt{-g} {\cal{L}}_{Dirac} d^5 x = \int  \sqrt{-g}\hspace{.03in}(i \bar{ \Psi} \Gamma^{a} {\cal{D}}_{a} \Psi) d^5 x
\ee

 where $g = \mbox{det}(g_{ab})$ = the determinant of five dimensional metric. 
The matrices $\Gamma^{a} = (e^{f(\sigma)} \gamma^{\mu}, -i \gamma^{5})$ provide a 
four dimensional representation of the Dirac matrices in five dimensional curved space. In this case $\gamma^{\mu}$ and $\gamma^{5}$ are the usual four dimensional Dirac matrices in chiral representation. The covariant derivative in 5D curved space for the metric in (\ref{metric}):

\begin{equation}
{\cal{D}}_{\mu} =(\partial_{\mu} -\frac{1}{2} f'(\sigma) e^{-f(\sigma)} \Gamma_{\mu}
\Gamma^{5}) ; \hspace{.4cm}
{\cal{D}}_{5} = \partial_{\sigma}
\end{equation}

The Dirac Lagrangian in 5D curved spacetime reduces to the following form
 
\begin{equation}
(-g)^{\frac{1}{2}} {\cal{L}}_{\small{Dirac}} = e^{- 4 f} \bar{\Psi}\left[i e^{f} \gamma^{\mu}\partial_{\mu} + 
\gamma^{5} (\partial_{5} -2 f') \right ] 
\Psi  
\end{equation}

The dimensional reduction of the action from five to four dimensions
 is performed in such a way that the standard four dimensional chiral 
particle theory is recovered. Since the four dimensional massive
fermions require both the left and right chiralities
it is convenient to organise the spinors with respect to
$\Psi_{L}$ and $\Psi_{R}$ which represent four component spinors living in five
dimensions given by $\Psi_{L,R} = \frac{1}{2} (1 \mp \gamma_{5})\Psi$.
Hence the full 5D spinor can be split up in the following way

\begin{equation}
\Psi(x^{\mu},\sigma) = \left( \Psi_{L}(x^{\mu})\xi_{L}(\sigma) +
\Psi_{R}(x^{\mu})\xi_{R}(\sigma) \right)
\end{equation}
where $\xi_{L/R}(\sigma)$ satisfy the following eigenvalue equations:

\begin{eqnarray}
\label{fermieq1}
 e^{-f(\sigma)}\left [\partial_{\sigma}-2 f'(\sigma) \right ] \xi_{R}(\sigma)
& = & -m~ \xi_{L}(\sigma) \\
 e^{-f(\sigma)} \left [\partial_{\sigma}-2 f'(\sigma)  \right ] \xi_{L}(\sigma)
& = & m ~\xi_{R}(\sigma)
\label{fermieq2}
\end{eqnarray}

The mass of the four dimensional fermions is denoted by $m$.
The full 5D action reduces to the standard four dimensional action 
for the massive chiral fermions, 
when integrated over the extra dimension, if
(i) the above equations are satisfied by the bulk 
fermions and (ii) the following orthonormality conditions are obeyed.

\begin{eqnarray}
\label{fermicond}
\int_{-\infty}^{\infty} e^{-3 f} \xi_{L_{m}} \xi_{L_{n}} d\sigma &=&
\int_{-\infty}^{\infty} e^{-3 f} \xi_{R_{m}} \xi_{R_{n}} d\sigma = \delta_{m n} \nonumber \\ 
\int_{-\infty}^{\infty} e^{-3 f} \xi_{L_{m}} \xi_{R_{n}} d\sigma &=& 0
\end{eqnarray}

Therefore, to achieve
localized modes one requires the extra dimensional part of the field to peak 
around the brane and the full solution to be normalizable and finite everywhere.
Following the usual technique of dimensional reduction we find the localisation conditions. If the function is sharply peaked around the brane we consider the field to be localised. On the other hand for a growing function away from the brane the field is not localised.

We now focus on the general mechanism for gravity localization on the brane. Finding the  effect of small gravitational fluctuations we study the generic properties
of gravity in the braneworlds. 
We investigate the metric perturbations restricted to the four dimensions around the classical solutions of the warped spacetime (\ref{metric}). The metric fluctuation can be written as 

\be
ds^2 = d\sigma^2 +e^{-2f(\sigma)} (\eta_{\mu\nu} + 
h_{\mu\nu}) dx^{\mu} dx^{\nu} \label{metricfluc}
\ee

where, {$h_{\mu\nu}(x^{\mu},\sigma) \sim \psi(\sigma) ~\hat{h}_{\mu\nu} (x^{\mu})$}. The four dimensional part obeys the canonical equation of motion: {{$\Box_{x}
\hat{h}_{\mu\nu} = m^2 \hat{h}_{\mu\nu}$}}. The mass of 4D graviton is given by $m$. 
 Let us restrict ourselves to study the
localization of the transverse traceless modes of the 
gravitational fluctuations which represent the 
gravitons on the brane. The transverse traceless (TT) modes are 
projected out by assuming $\partial_{\mu} h^{\mu}_{\nu} = 0$ and $h^{\mu}_{\mu} 
= 0$.  The variation of the Einstein tensor gives us the equation of motion for the higher dimensional part of the  graviton as \cite{gravloc, csaki}

\be
\label{gravieq1}
\frac{1}{\sqrt{-g}} \partial_{\sigma} \left( \sqrt{-g} \partial^{\sigma} \psi(\sigma) \right) = m^2 e^{2f(\sigma)} \psi(\sigma)
\ee

The above equation always admits a solution $\psi(\sigma) = const.$ when
the graviton is massless. This leads to the zero modes of the gravity 
fluctuation. The fluctuation equation (\ref{gravieq1}) can be
derived from the canonical form of the action
\bea
S & \sim & \int d^{5}x \sqrt{-g} \partial_{R}h_{MN} \partial^{R} h^{MN} \nonumber
\\
{} &=& {} \int d\sigma g^{00}(\sigma) \sqrt{-g(\sigma)} \psi(\sigma)^2 \cdot 
\int d^{4}x \partial_{\rho} \hat{h}_{\mu\nu} \partial^{\rho}
\hat{h}^{\mu\nu} + \cdots
\eea  

Performing the dimensional reduction from 5D to 4D we obtain the localization condition for gravity as 

\be
\label{gravnorm}
\int \sqrt{-g} g^{00}(\sigma) \psi_m \psi_{m'} d \sigma = \delta_{mm'}
\ee

Following similar mechanism one can easily obtain the localisation condition for the scalar fileds on the brane \cite{5dloc}. Let us explore the localisation scenario in different kinds of brane world models in the following sections.

\section{Fermion localisation}


The extra dimension is infinitely extended in the RS-II model. The warp factor is a decaying function of $\sigma$. It is clear from the localization conditions given in (\ref{fermicond}) that the massless fermions are not localized on the brane. It has been already stated that one can obtain massless fermions on the RS-II brane by considering additional interaction with bulk fields \cite{fermiloc}. In the following we will study the role of bulk scalars in the geometry as well as in the localization of fermions on the brane.  

Let us consider a minimally coupled real scalar field with sine-Gordon (SG) potential \cite{rksk1}. The exact solution for the warp factor is given by :
\be
e^{-2f(\sigma)} = A \cosh^{- 2 \nu} (b \sigma) \label{sg}
\ee
where $\nu$, $A$ and $b$ are constants dependent on the parameters of the theory. The scalar field is a soliton. This model with a bulk SG potential provides a `thick brane' realisation of the Randall--Sundrum scenario where the SG field and its soliton configuration dynamically generate this domain wall configuration
in the background warped geometry. In addition, as is obvious from the
functional from of $f$, there is no discontinuity in the derivative of
$f$ at the location of the brane. The warp factor is smooth everywhere.

Similarly for a  bulk phantom scalar field (with negative kinetic energy) with
sine-Gordon potential one can generate an exact thick brane solution with an increasing warp factor given by

\be
e^{-2f(\sigma)} = B \cosh^{2 \rho} (c \sigma) \label{phantom}
\ee
where $\rho$, $B$ and $c$ are constants dependent on the parameters of the theory. This also provides a `thick brane' model where the SG field and its soliton
configuration dynamically generate this domain wall configuration in the warped geometry\cite{rksk2}.
We further consider a non minimally coupled scalar field. The action for the bulk scalar field {\cite{rksk3}} given by 

\begin{equation}
S_{T} = \alpha_{T} \int d^{5}x \sqrt{-g} V(T)\sqrt{1+g^{ab}\partial_{a} T\partial_{b} T}
\end{equation}

where $\alpha_{T}$ is an arbitrary constant, $g_{ab}$ being the
five dimensional metric. The scalar field is represented by T and 
V(T) corresponds to its potential. The constant $\alpha_{T}$ can take
either positive or negative values. 
For a chosen form of the potential we obtain an exact solution for the warp factor as 
\be
e^{-2f(\sigma)} = e^{\frac{2 a_1}{k}e^{- k \vert \sigma \vert}} \label{tachyon}
\ee
where $a_1$ is a constant. It is important to note that
the metric warp factor is a super exponential function which decays
as one moves along the transverse dimension. The scalar is an exponentially
growing function of $\sigma$. In this case the scalar
field never becomes zero, keeping the potential always
non-singular {\cite{rksk3}}. 

 As mentioned earlier, one can achieve localized fermions on the brane by
introducing the Yukawa coupling between fermion field and the scalar
field, $\eta_{F} \bar{\Psi} \mbox{F}(\Phi) \Psi$ where
$\mbox{F}(\Phi)$ is a function of the scalar field and $\eta_{F}$ is
the coupling constant. The Lagrangian for a Dirac fermion now becomes
\be
\label{coupledaction}
\sqrt{-g} {\cal{L}}_{Dirac} = \sqrt{-g}\hspace{.03in}(i \bar{ \Psi} \Gamma^{a} {\cal{D}}_{a} \Psi
- \eta_{F} \bar{ \Psi} \mbox{F}(\Phi) \Psi )
\ee
The Yukawa coupling between the scalar and the fermion, 
with the kink solution for the scalar,
is necessarily like an effective, variable, 5D mass 
for the fermions \cite{fermiloc}. This is largely responsible for
generating the massive fermion modes in four 
dimensions. The dynamical features of the models 
can thus be obtained from the solutions of the eigenvalue equations 
(\ref{fermieq1}) and (\ref{fermieq2}). In this paper we only address the localization of massless fermions. Although the massive fermions can also be found on the brane by this mechanism \cite{rksk1, rksk2}. 

Let us first focus on massless (i.e. $m = 0$) fermions for a Yukawa coupling of the form
$\eta_{F} \bar{\Psi} \Phi \Psi$ for the metric (\ref{sg}). In this case the equations (\ref{fermieq1}) and (\ref{fermieq2}) reduce into two decoupled equations. The solutions of which are asymptotically written as
\begin{eqnarray}
\xi_{L}(\sigma) & = & e^{-\left (\frac{\eta_{F}\pi}{2}  - 
2 \sqrt{\frac{\vert \Lambda \vert}{6}} \right ) \vert \sigma \vert} \label{sgzerol}\\
\xi_{R}(\sigma) & = & e^{\left (\frac{\eta_{F} \pi}{2}  + 
2 \sqrt{\frac{\vert \Lambda \vert}{6}} \right ) \vert \sigma \vert} \label{sgzeror}
\end{eqnarray}
where $\Lambda$ is bulk cosmological constant. Eqn. (\ref{sgzerol}) yields  the localization of left chiral fermions on the brane so long as,
$\eta_{F} \ >  \frac{4}{\pi} \sqrt{\frac {\vert \Lambda \vert}{6}}$ for a kink profile of the bulk scalar.
One can as well achieve right chiral states by considering the interaction with an anti-kink sine-Gordon profile. 

Let us now focus our attention on localization of spinor fields on the
brane with an increasing warp factor (\ref{phantom}) in the background of a bulk
phantom scalar field. The coupling of bulk fermion with the scalar field 
gives rise to two chiral fermionic zero modes in four dimensions. 
Depending on the sign of the coupling constant one of these two modes is found to be localized on the brane while the other is delocalized and not normalizable \cite{fermiloc}.  Generally the Yukawa coupling with the bulk scalar 
is capable of localizing only a single chiral state (right or left) while
fermions of both chiralities are expected. 
For the warped geometry given by the metric (\ref{phantom}) the eigenvalue
equations (\ref{fermieq1}) and (\ref{fermieq2}) have the
asymptotic solutions of the following form 
\begin{eqnarray}
\xi_{L}(\sigma) & \sim & e^{-\left (\eta_{F}  + 2 \sqrt{\frac{\vert \Lambda \vert}{6}}
\right ) \vert \sigma \vert} \\
\xi_{R}(\sigma) & \sim & e^{ \left (\eta_{F}  - 2 \sqrt{\frac{\vert \Lambda \vert}{6}}
\right ) \vert \sigma \vert}
\end{eqnarray}
We found that in the background
geometry with an increasing warp factor both the left and right chiral 
zero modes are confined to the brane even without any coupling with the bulk field. 
The novel picture is that both the left and right chiral massless modes are normalizable even for $\eta_{F} = 0$, which is in accord with the statements on exponentially
rising warp factors in {\cite{bajc}}. 

Similarly, for the warped geometry given by the metric (\ref{tachyon}) the massless modes of the fermions are given by
\begin{eqnarray}
\xi_{L} = exp \left[ \frac{2}{k} \left(- a_{1} e^{- k \vert \sigma \vert} - \eta_{F}
  \sqrt{\frac{2}{a_{1} k}} e^{\frac{k}{2} \vert \sigma \vert} \right) \right]
  \\
\xi_{R} = exp \left[ \frac{2}{k} \left(- a_{1} e^{- k \vert \sigma \vert} + \eta_{F}
  \sqrt{\frac{2}{a_{1} k}} e^{\frac{k}{2} \vert \sigma \vert} \right) \right] 
\end{eqnarray}

If the coupling constant, $\eta_{F} = 0$, it is clear from the above expressions that both
the left and right chiral modes decay away from the brane. If
$\eta_{F} > 0$, then $\xi_{L}$ decays but $\xi_{R}$ grows and the
reverse phenomena takes place for $\eta_{F} < 0$.

 An interesting feature has been observed in the multiply warped spacetime introduced in \cite{dcssg}. The bulk geometry is formed by the warped compact dimensions which get further warped by a series of successive warping leading to 
multiply warped spacetime with various p-branes sitting at the different 
orbifold fixed points. The geometry of the multiply warped D dimensional spacetime is given by : $M^{1, D-1} \rightarrow \left\{ [M^{1,3} \times S^1/Z_2] \times S^1/Z_2 \right\} \times \cdots$, with $(D - 4)$ such warped directions. For a negative cosmological constant in the bulk one can find a doubly warped spacetime given by the metric: 
\be
ds^2_6 = \frac{\cosh^2(kz)}{\cosh^2(k \pi)}  \left[ \exp(- 2 c \vert y \vert) \eta_{\mu\nu}dx^{\mu}dx^{\nu} + R^2_y d y^2 \right ] + r^2_z d z^2
\ee
where the orbifolded compact directions are denoted by the angular
coordinates $y$ and $z$ respectively with $R_y$ and $r_z$ as respective modulus.
The most important feature in the multiply warped scenario is that 
the $Z_2$ orbifoldings gives rise to coordinate-dependent
brane tensions on two 4-branes located at $y = 0$ and $y = \pi$ \cite{dcssg}. These are given by  
\be
V_{y = 0} = - V_{y = \pi} = V_{0} ~\sech~(kz)
\ee
In this scenario the coordinate dependent brane tension effectively plays the role of a bulk scalar field which appears naturally from the requirement of orbifolded boundary conditions along the two internal compact directions. The 3-branes are located at the edges of the 4-branes on locations $y = 0, z =0$, $y = 0, z = \pi$, $y = \pi, z =0$ and $y = \pi, z = \pi$. We consider the five dimensional fermions residing on the 4-branes at $y = \pi$.  In this scenario the coupling between the 5-dimensional fermions and the brane tension (i.e. the scalar field distribution) plays an important role in localization of fermions on the 3-branes. 
For appropriate choice of the coupling parameter between the 
5-dimensional fermions and the scalar field distribution only the left chiral mode of 
the fermion can be localized on our TeV brane while the right handed mode gets more and more localized towards the other 3-brane lying at the other edge of the wall. In the figures (\ref{multiloc1}) and (\ref{multiloc2}) we have plotted the left and right chiral modes for different coupling strengths.

\bfig[ht]
\includegraphics[height = 5.5cm, width = 8.5cm]{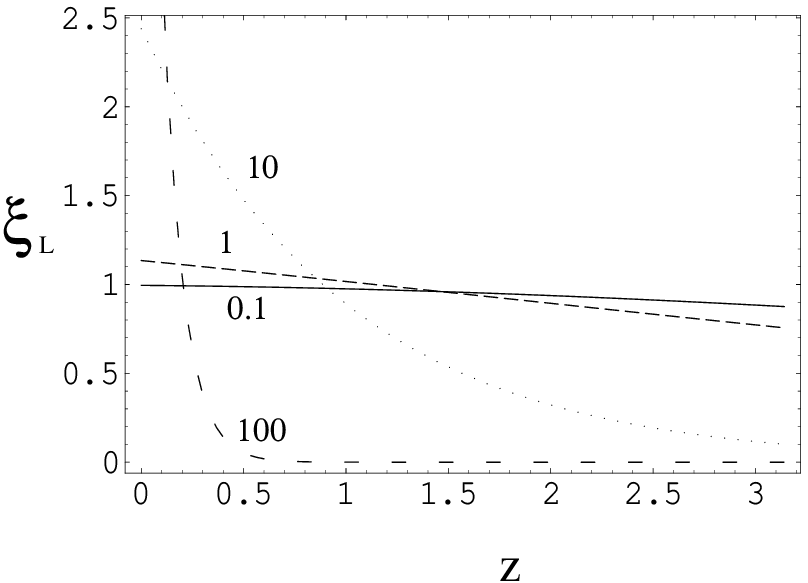}
\caption[zeromode]{We have plotted the left chiral modes for several values of $\eta$.}
\label{multiloc1}
\efig

\bfig[htb]
\includegraphics[height = 5cm, width = 8cm]{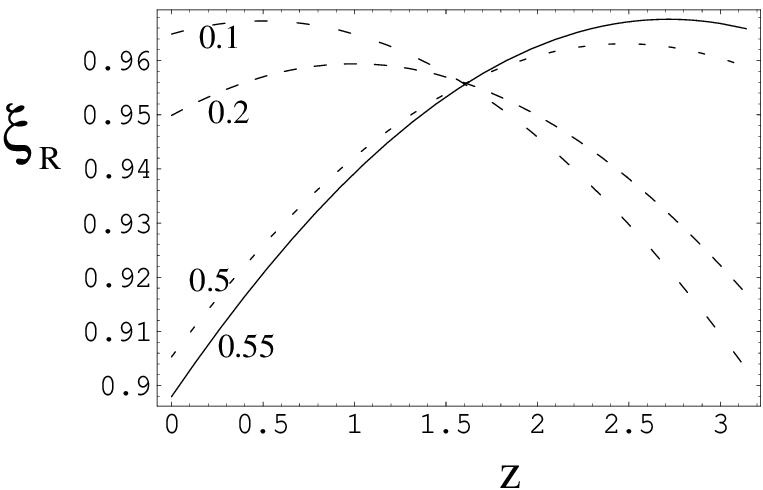}
\caption{Right chiral modes are plotted for different values of $\eta$.}
\label{multiloc2}
\efig 

This phenomenon thus offers a natural explanation of the origin of chiral massless fermion mode with only one chiral component in our 3-brane. Thus one does not need to invoke some external scalar field by hand to achieve the localization. The consistency requirement of the theory itself provides a mechanism for chirality preferential localization \cite{rkssg}.

\section{Gravity Localisation}

The obvious question -{\em{ how gravity should behave if we have extra
dimensions}} - has been addressed quite ingeniously by Randall and Sundrum
in the so called RS-II model {\cite{RS}}. They have shown by using a non-factorizable ``warped'' geometry for the extra dimension, one could in fact have an infinitely large extra dimension and still reproduce Newton's law at large distances on the brane. The key observation of RS is that in their scenario there is a localized zero-energy graviton bound state in 5D which should be interpreted as the ordinary 4D graviton. The correction term to the Newton's law comes from the KK tower of gravitons. The gravitational potential between to test particles of mass $m_1$ and $m_2$ is given by 
\be
 V(r) = G_N \frac{m_1 m_2}{r} \left( 1 + \frac{1}{r^2 k^2}\right)
\ee

We now investigate the graviton zero modes for the braneworld models with bulk scalars discussed in the previous section. 
The warp factor due to a bulk scalar with sine-Gordon potential has an explicit dependence on the parameter $\nu$ (Eq. (\ref{sg})).  The
equation governing the behavior of $\psi(\sigma)$, Eq. (\ref{gravieq1}), reduces to a
analogous quantum mechanics problem for the parameter $\nu
=1$. First, we will study the localization scenario for this particular
choice of the parameter and later discuss about the issues in general.    
Applying the normalization condition (Eq.  (\ref{gravnorm})) to the metric
we obtain localized zero modes on the brane.  
Let us perform a coordinate transformation $z = \frac{1}{b} \sinh (b \sigma)$ to obtain the perturbed metric in a conformally flat form
\begin{equation}
ds^2 = e^{- 2 f(z)} [(\eta_{\mu\nu} + h_{\mu\nu})dx^{\mu} dx^{\nu} + dz^2 ]
\end{equation} 
\noindent where $f(z) = \frac{1}{2} \log (1 + b^2 z^2)$ 
and $b =\frac{\kappa_{1}}{\alpha}$. Redefinition of the z dependent
part of each component of the metric perturbation as 
$H(z) = e^{-3 f(z)/ 2} \psi(z)$ reduces the graviton equation into a 
Schr\"{o}dinger equation for a quantum mechanics problem 
\begin{equation}
[- \partial_{z}^2 + U(z)]H(z) = m^2 H(z) \label{modeeq} 
\end{equation}
where, the potential is given by 
\begin{equation}
U(z) = -\frac{3}{2} \frac{b^2}{(1 + b^2 z^2)} + \frac{21}{4} \frac{b^4 z^2}{(1 + b^2 z^2)^2}
\end{equation}

\begin{figure}[h]
\centerline{
\includegraphics[width= 8cm,height=5.1cm]{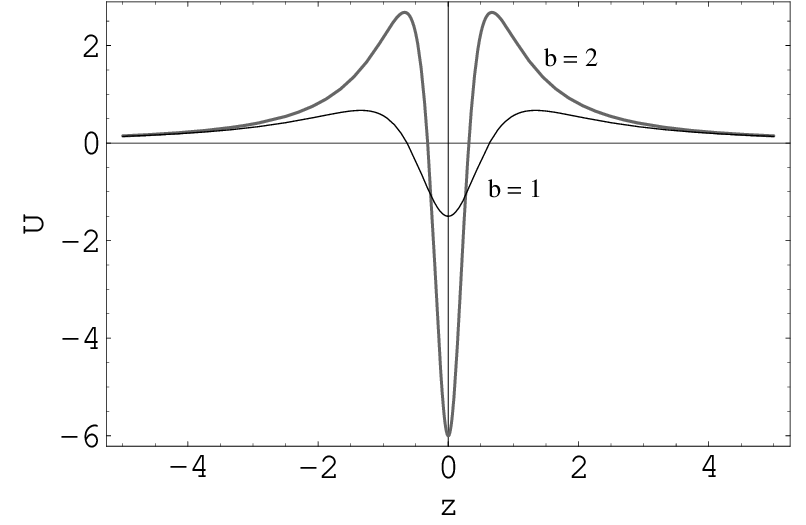}}
\caption{The volcano potential for gravitational fluctuations plotted
as a function of conformal coordinate z.} 
\label{gravpot}
\end{figure}

The potential U(z) is plotted in Fig. (\ref{gravpot}) for two different values
of the parameter b.
Larger b value corresponds to deeper minimum for the potential. This is the same type of
volcano potential as obtained in {\sc rs-ii}. It is important that the
potential falls off asymptotically slower than that in the AdS case. 
This in turn ensures the existence of normalizable states for the 
wave function of $m = 0$ eigenvalue \cite{csaki}. Here, $U(z) = 0$ as 
$\vert z \vert\rightarrow \infty$. This is much more
interesting compared to the cases where U(z) is either asymptotically
positive or negative as in the former case massless zero modes will be
always normalizable and in the other case not at all normalizable.
The zero modes ($ m^2 = 0$ ) can be represented as 
\begin{equation}
H(z) = N (1 + b^2 z^2)^{-\frac{3}{4}}
\end{equation}
where, $N$ is the normalization constant. The localization 
of the massless modes can be ensured from the finiteness of the
following integral 
\begin{equation}
\int H^2 dz < \infty
\end{equation}
The factor f(z) tends to infinity at a sufficiently fast rate for
$\vert z \vert \rightarrow \infty$. As a consequence, the massless
graviton mode falls off rapidly as one moves away from
the brane which in turn reproduce Newtonian gravity on the brane.
\begin{figure}[h]
\centerline{
\includegraphics[width= 8cm,height=5.5cm]{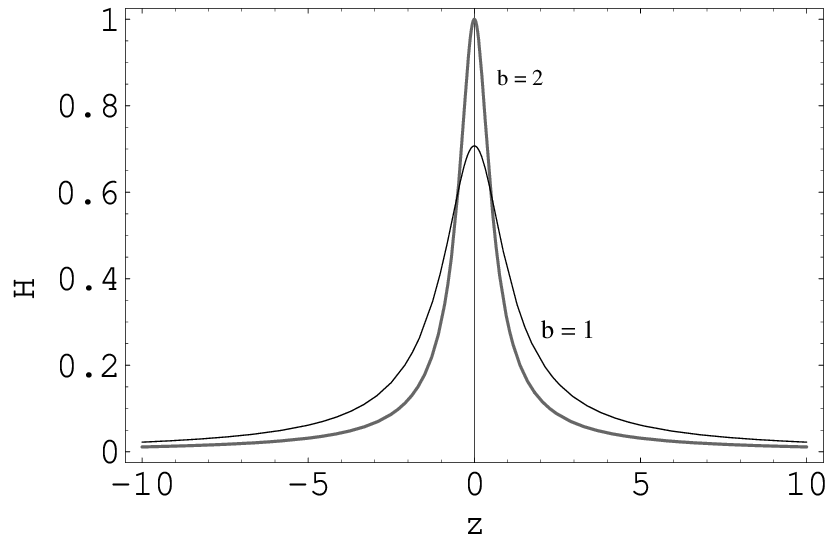}}
\caption{Zero modes of the gravitational fluctuations plotted
as a function of conformal coordinate.} 
\label{gravzero}
\end{figure}

In Fig. (\ref{gravzero}), we have shown the massless modes for the same values of b
as used in the potential. There is a sharp peak for the larger b value
which implies that the probability of getting bound states on the
brane gets more pronounced for a deeper potential at the location of
the brane. It follows from the nature of the warp factor that 
for the metric due to a bulk phantom field with sine-Gordon potential
(Eq. (\ref{phantom}), massless gravitons are not localized on 
the brane. On the other hand, in the thin brane model for a bulk
tachyon field (Eq. (\ref{tachyon})) the graviton zero modes
are confined to the brane.

The localization scenario discussed so far in this paper and also
in the work of others \cite{5dloc, fermiloc, gravloc}, reveal that all the standard model fields
 are not localized on a single brane only through gravitational 
interactions. In search of such a braneworld model one may introduce
another extra dimension. This is our course of study in the next
section where we will see that it is possible to localize 
several standard model fields as well
as gravity  on a single brane embedded in a six
dimensional bulk spacetime.


\section{Localization of all fields}
Let us search for a braneworld model which can localize all the SM fields as well as gravity on a single brane we consider two extra dimensions. For instance, in \cite{ 6dloc} it
was claimed that all the zero modes of the {\sc sm} fields can be
localized on a single brane by means of the gravitational interaction only. 
Let us begin with the following metric ansatz for a warped
brane embedded in six dimensions: 
\be
ds^2 = e^{2f(r)} \eta_{\mu\nu}dx^{\mu}dx^{\nu} + d r^2 + e^{2 g(r)} L^2 d\theta^2 
\ee
where the radial coordinate r is infinitely extended ($0 \le r < \infty$) 
and the compact coordinate $\theta$ ranges from $0 \leq \theta \leq 2\pi$. L is 
an additional parameter characterizing the extra compact direction
on the 4--brane. We also assume that the warp factor $f(r)$ and the function 
$g(r)$ depend on the extra dimensional radial coordinate, r, only. We now focus on  time-independent solutions of the Einstein equations for a bulk phantom scalar. The solutions for the warp factors are found to be :
\be
\label{6dmetric}
 ds^2 = e^{\frac{k}{2r}} \eta_{\mu\nu}dx^{\mu}dx^{\nu} + d r^2 + e^{- 2kr} L^2 d\theta^2 
\ee
where, k is an arbitrary constant. The warp factors show a distinct nature - the brane part is a growing function of r and the other part is a decaying function. In order to 
study the confinement of fields we employ the simplest test originally outlined in {\cite{bajc}}. For different spin fields we assume at the outset 
that they are not dependent on the extra coordinates. The consistency
check is then done by showing that the effective coupling constants 
emerging after dimensional reduction are non-vanishing and finite. To achieve
localized modes one requires the extra dimensional part of the field to peak 
around the brane and the full solution to be normalizable and finite everywhere.
Following the usual technique of dimensional reduction we find the localisation conditions. In this minimal set up the localization conditions for the massless modes of gravity, fermion and gauge fields are given by the equations (\ref{spin02}), (\ref{spin1/2}) and (\ref{spin1}) respectively.
\bea
\int^{\infty}_0 e^{2f(r) + g(r)} \psi_m \psi_{m'} dr = \delta_{mm'} \label{spin02} \\
\int^{\infty}_0 e^{-f(r)} \xi_m \xi_{n} dr = \delta_{mn} \label{spin1/2} \\
\int^{\infty}_0 e^{g(r)} \phi_m \phi_{n} dr = \delta_{mn} \label{spin1}
\eea 
where $\psi$, $\xi$ and $\phi$ represent the higher dimensional part of the fields.
It is miraculous to note that in the geometry given by the metric (\ref{6dmetric}) all the fields are localised on a single by gravitational interactions only \cite{rksk4}.

\section{Conclusion}

Let us now summarize the results presented in this paper :

We have a given the general mechanism of localisation of fields on a brane. Then the localisation scenario in the RS model model has been discussed briefly. The effect of bulk scalar fields on the warping of geometry has been discussed in detail. The  
fermion fields have been studied in the background geometry of
the three different kinds of bulk scalar fields. We couple the scalar field
to the spinor field through a Yukawa coupling and study the behaviour of massless modes in  thick brane models. Both the left and right chiral modes are found to be normalizable on the brane in the background geometry of a bulk phantom scalar field. 
Localization of gravity has been studied in case
of the braneworld model with a bulk scalar of sine-Gordon
potential. The massless modes are normalizable on the brane. 
We have also studied the localization of massless gravitons in 
the model with a bulk tachyon scalar.
In the exact background geometry obtained from a phantom field in the six dimensional bulk it is possible to have the zero modes of several standard model fields and gravity to be localized on the brane. The models involves four branes with an
on--brane compact extra dimension (hybrid compactification).

\end{document}